# How and When Did Locality Become 'Local Realism'? A Historical and Critical Analysis (1963-1978)

Federico Laudisa

Department of Philosophy and Humanities, University of Trento
Via Tommaso Gar 14, 38122, Trento, Italy

*federico.laudisa@unitn.it*

**Abstract**

The history of the debates on the foundational implications of the Bell non-locality theorem displayed very soon a tendency to put the theorem in a perspective that was not entirely motivated by its very assumptions, in particular in term of a 'local-realistic' narrative, according to which a major target of the theorem would be the very possibility to conceive quantum theory as a theory concerning 'real' stuff in the world out-there. I present here a historico-critical analysis of the stages, between 1963 and 1978, through which the locality condition of the original Bell theorem almost undiscernibly turned into a 'local realism' condition, a circumstance which too often has affected the analysis of how serious the consequences of the Bell theorem turn out to be. In particular, the analysis puts into focus the interpretive oscillations and inconsistencies that emerge in the very descriptions that many leading figures provided themselves of the deep work they devoted to the theorem and its consequences.

# 1  Introduction

In a lecture held in 1983, Richard Feynman argued that the Bell theorem "is not a theorem that anybody thinks is of any particular importance. We who use quantum mechanics have been using it all the time. It is not an important theorem. It is simply a statement of something that we know is true – a mathematical proof of it." (quoted in Whitaker 2016b, 493): in Feynman's view, what 'we know is true' is simply that quantum theory is not a classical theory. No matter what is the tenability of the Feynman charge of irrelevance about the Bell theorem, a common view of what it takes for a physical theory to be 'classical' is that the physical systems the theory is about can be assumed to have measurement-independent properties or, in other terms, that – in the well-specified situations that are suitable for physical investigation – these physical systems can be assumed to have *pre-existing values* for all relevant quantities, values that the measurement is supposed just to reveal. In this vein, 'classicality' is thus equated more often than not with a loose notion of 'realism'. As the following formulation, due to Chris Isham, describes effectively, in this framework neither the notion of 'pre-existence' of values, nor the notion of 'revealing' measurement are problematic:

> The notion of an object as the bearer of determinative properties lies at the heart of realism, which is one of the main Western, 'commonsense' philosophical positions. […] The epistemological question of how we can have knowledge of the properties of an object is answered in physics by the notion of *measurement*, i.e., any physical operation by which the value of a physical quantity can be determined (perhaps only to a certain accuracy) and recorded. Such a picture is in accord with the general object-subject split of scientific methodology whereby part of the natural world is deliberately isolated from its environment so that theoretical and experimental investigations may proceed unhampered by any influence from the rest of the world. From the perspective of classical physics, the separation of observer and system has no fundamental significance. Observer and observed are both part of a single, objectively-existing world in which, ontologically speaking, both have equal status, and are potentially describable by the same physical laws. Similarly, there is nothing special about the concepts of 'measurement' or 'observable'. The reason why the measurement of an observable quantity yields one value rather than another is simply because the quantity *has* that value at the time the measurement is made. Thus properties are intrinsically attached to the object as it exists in the world, and the measurement is nothing more than a particular type of physical interaction designed to display the value of a specific quantity. (Isham 1995, p. 68-69).

That this is not the case with quantum mechanics is already at the heart of the foundational debates, surrounding the establishment of quantum theory in the first thirty years of the twentieth-century[1], but the well-known results by A.M. Gleason, S. Kochen, E. Specker and J.S. Bell between 1957 and 1967 provided this feature of quantum theory with a rigorous and unequivocal foundation. When the 1952 Bohm alternative formulation of quantum mechanics came into focus in the 60's, after a decade of virtual disregard by the scientific community, the time could have been taken to be ripe, then, for restricting the attention to the real issue on which also the Bohm theory was challenging a commonsense view of physical reality, namely non-locality. And, in fact, the first formulation of the Bell theorem in 1964 was exactly the outcome of an analysis according to which *any* theory, in whatever formulation, could recover the statistical predictions of quantum mechanics only at the condition of being non-local.

Very soon, however, attempts to deflate this revolutionary impact of the theorem started to develop and, with different tools and aims, still continue today. The first of these attempts has been the reading of the Bell theorem simply as a no-hidden-variables result, a reading that is blatantly contradicted by the very existence of the above mentioned Bohm version of quantum mechanics, now commonly called Bohmian mechanics[2]. According to a more recent and refined version of this deflationary attitude, the alleged real target of the theorem includes a (vague) 'realism' along the lines sketched above: such target, that is, would be a conjunction of locality *and* this 'realism' and this is what after all justifies the label 'local realism' for the worldview that – in *this* reading – the Bell theorem would make untenable. This attitude has become the mainstream view of the state of affairs in the area of the investigations on the consequences of the Bell theorem, and expressions of it are widepread in the literature. According to Pawlowski and Brukner,

> quantum theory predicts correlations between spacelike separated events, which are nonsignaling but cannot be explained within *local realism*, i.e., within the framework in which all outcomes have preexisting values for any possible measurement before the measurements are made («realism») and where these values are independent from any action at spacelike separated regions («locality»)" (Pawlowski, Brukner 2009, p. 030403-2, my emphasis) whereas, according to Wiseman,

> the world is made up of real stuff, existing in space and changing only through local interactions – this *local-realism* hypothesis is about the most intuitive scientific postulate imaginable. But QM implies that it is false." (Wiseman 2015, p. 649, my emphasis).

---

[1] Two references, out of a myriad of possible examples, can be Heisenberg 1930 and Bohr 1939.
[2] As is well-known, Bell himself disliked the 'hidden-variables' jargon and (rightly) considered the hidden-variables terminology seriously misleading.

The claim that what the Bell theorem is about is the refutation of local realism has acquired the status of a commonplace, to such an extent that any new experimental confirmation of the nonlocal behaviour of entangled states – also in contexts inspired by quantum technology which are not interested into the philosophical side of the issue – is advertized as a blow to 'local realism'. In their proponents, this attempt to resist the radical implication of Bell's nonlocality for our ordinary spacetime theories has a clear advantage: the possibility to preserve locality, by discharging the import of the theorem on the side of 'realism'. Given the anti-realistic folklore surrounding quantum mechanics anyway since the inception of the theory, this move comes at a relatively little price and contributes to downplay significantly the foundational relevance of Bell's theorem.

On the other hand, there have been several works that have shown in a rigorous and convincing way that this 'local-realistic' reading is misplaced and I will side with them, taking for granted that such reading fails to be correct beyond any reasonable doubt (Norsen 2007, XXX 2008, 2012, 2017, 2018, Maudlin 2014). This position motivates the proposal that I put forward in the present paper: a historico-critical analysis of the stages through which the locality condition almost undiscernibly turned into a 'local realism' condition, a circumstance which too often has affected the analysis of how serious the consequences of the Bell theorem turn out to be. How did the consensus on 'local realism' get established, in spite of its dubious logical status? Which were its underying motivations and how and why did this dubious logical status fail to be adequately recognized? My analysis will cover the span 1963-1978, with the Jauch-Piron 1963 paper – a major pre-Bell 1964 contribution – as a starting point (Jauch, Piron 1963), through important papers such as Clauser, Horne, Shimony, Holt 1969 and Clauser, Holt 1974 up to the important review paper by Clauser and Shimony in 1978. The present work aims to clarify the history of this misinterpretation narrative, contributing to dispel the confusion that still is present when most of the main reference papers between the 60's and the 70's are reconstructed. In section 2, I analyze the role of 'realism' *before* the Bell theorem by locating an instance of it in the 'no-hidden-variables' theorem proved by Jauch and Piron in their 1963 paper: I argue that the motivation for an assumption which is crucial for this no-go theorem can be interpreted along 'realistic' lines, a circumstance that seriously restricts the scope and relevance of the Jauch and Piron result. Interestingly, this motivation is distinct from the more familiar objection raised with respect to the Jauch-Piron result by Bell himself in his 1966 paper (written, in fact, immediately after the publication of the Jauch-Piron paper). In section 3, I focus on the 1964 Bell celebrated paper by showing how such paper starts exactly where the EPR argument ends: namely, from a condition – represented in the EPR paper by the condition for Elements of Physical Reality – which states essentially what EPR *mean* by an objective property and which does *not* assume at the outset the pre-existence of such properties. In section 4, I focus on the 1969 Clauser-Horne-Shimony-Holt paper, which is meant to be the first experimentally meaningful generalization of the Bell theorem, and I emphasize that a close analysis of the paper shows the complete irrelevance of any 'realistic' assumption in deriving the CHSH inequality. Nevertheless, this is the place where this Locality-to-'Local Realism' transition starts to emerge, a transition that appears to be completed and undisputable already in the 1974 Clauser-Horne paper and the 1978 Clauser Shimony paper, which are subject to analysis in section 5. Some concluding remarks are then formulated in the final section 6.

## 2      'Realism' *before* the Bell theorem (1963-1966)

As is now well known, John S. Bell was unsatisfied with the ordinary 'Copenhagen' school of thought well before his celebrated papers of the Sixties, but at the beginning of the 1950's he seemingly was not even aware of the EPR paper. According to a personal recollection of Leslie Kerr (a Bell mate in Queen's College in Belfast), "Bell had not seen it while in Belfast" and, according to the Bell biographer Andrew Whitaker, "it seems very likely that he met it for the first time in Bohm's book" (Whitaker 2016a, p. 104). The Bohm's book is of course the quantum mechanics textbook (Bohm 1951), which is known in the history of the debates on the foundations of quantum mechanics for two main reasons: first, it contains the first simplified formulation of the EPR argument in terms of spin; second, it was in fact a turning point for Bohm himself, who developed a critical attitude toward Copenhagen quantum mechanics via the very effort of thoroughly studying quantum mechanics in order to expose it in the textbook[3]. After his meeting with the EPR paper and the two 1952 papers in which Bohm proposed his full-fledged hidden variable formulation of quantum mechanics, though, Bell never stopped to engage with foundational issues as deep issues at the heart of real physics: in those papers he had seen 'the impossible done', as he vividly expressed himself in a much later paper (Bell 1982). His wife Mary recalled:

> When the David Bohm papers appeared in 1952, I can remember how excited he was. In his own words, 'the papers were for me a revelation'. When he had digested them, he gave a talk about them to the Theory Division. There were interruptions, of course, from Franz Mandl, with whom he had had many fierce arguments. (Bell M. 2002, pp. 4-5)[4].

---

[3] The book itself, in fact, was quite successful. As a Bohm biographer recalls, "upon publication, *Quantum Theory* gained favorable reviews and was adopted by a number of universities for their courses. Bohm sent copies to several leading physicists. Pauli replied with a warm letter approving of the way Bohm had woven mathematics and physics together and addressed philosophical questions." (Peat 1996, p. 109).

[4] It is not surprising what Bill Walkinshaw, a Bell's colleague, recalled: "when Bohm came to give a lecture, it was John who shone at question time and it was clear that he had studied Bohm's work with some care."

Bell was also in the audience of a lecture that Bohm gave at University College, London, on his own interpretation (called *causal* at that time) of quantum mechanics. Bohm himself recalled the episode in an interview:

> I gave a talk in University College on the Causal Interpretation including the spin, which was well received. John Bell was there. He was a young physicist who later went to Geneva and he later worked out this Bell's Theorem which people talk about it a lot on locality now. He was affected by this talk. He had believed that there was no way to understand quantum mechanics causally. Then suddenly he said, "Here it is in front of you." (Wilkins 1986).

These circumstances largely justify the Bell remark in the *Acknowledgements* of his 1966 paper, his *first* paper on hidden variables, according to which "the first ideas of this paper were conceived in 1952". There Bell actually mentions his discussions with just two colleagues: Fritz Mandl (mentioned above by Mary Bell) and Josef M. Jauch, the dialogue with whom will prove to be also quite relevant for setting the stage to the later non-locality result (we will soon return to Jauch)[5].

According to the above remark, something similar to the incubation of the Kant transcendental enterprise – from the 1770 *Dissertation* to the 1781 first edition of the *Critique of the Pure Reason* – had occurred to Bell: more than ten years separate his discovery of the Bohmian framework from the criticisms of a series of no-go theorems on hidden variables in his first paper on the topic. That such a long time of reflection elapsed without actually producing any paper on the subject was due, in the case of Bell, to a contingent circumstance: on leave from his job at the Atomic Energy Research Establishment at Harwell, in 1953 Bell studied in Birmingham under the supervision of Rudolf Peierls, a physicist who was not especially fond of foundational issues in quantum mechanics, issues that were taken up again when in 1960 Bell moved to CERN in Geneva. The attitude exemplified by Peierls was also the indication of a more general circumstance, concerning the status of fundamental physics and in particular the difference between the pre- and post-war state of affairs of scientific education in the wider physics community. In the decade which gave rise, among others, to the EPR paper and the Schrödinger cat paper, the

---

(Whitaker 2016a, p. 123). In an interview given in 1986, Bell said that the issue that gave rise to the derivation of his inequalities "has been at the back of my head continually", namely well before 1964.

[5] As is presently well-known (Whitaker 2016a, pp. 200-201), the publication procedure for this paper was unfortunate. Bell could certainly profit from a friendly review, since the journal chose David Bohm himself as a referee, but when he sent the final version of the paper back to the journal for publication, the paper was misfiled and some time elapsed before the managing editor wondered why he had received no reply from Bell. When finally the editor wrote Bell asking for explanation, the letter was sent to the address from where the original paper had been sent as a submission (at the SLAC laboratory), but Bell could not be found there anymore, since by that time he had returned to CERN in Geneva. Only much later Bell himself and the editor realized what had happened, so that the paper finally found its way to publication only in 1966.

philosophical and foundational issues were deeply intertwined with the strictly physical issues, and many physicists involved in the debate, no matter what their attitude was toward general questions like the very meaning of the quantum theory, still agreed that an advance in the former could provide an advance in the latter. As the historian of physics David Kaiser nicely puts it:

> During the 1920s and 1930s the architects of quantum mechanics – most of whom were European – tackled the deep philosophical implications of the subject in their classrooms and textbooks head on. No clean line separated calculation from interpretation. Theorists like Bohr, Heisenberg, Hermann Weyl, Max Born and Arnold Sommerfeld each paused within their textbooks to relate the latest discoveries in atomic physics to long-standing trends in philosophical inquiry, such as the extent to which we can ever gain trustworthy knowledge about the physical world (let alone unobservable entities), the role of language in shaping our concepts, or the active filtering of seemingly direct observations by our prior concepts such as space and time. Some invoked the philosophers Immanuel Kant or Ernst Mach; others even turned to Eastern mysticism and Jungian psychoanalysis to help interpret the latest physics research. *But all agreed that the new physics demanded serious philosophical scrutiny*. (Kaiser 2007, p. 30, my emphasis)

Things were bound to change dramatically starting from the 1940's. Not only the successes of the extension of quantum mechanics to quantum field theory obscured the relevance of deciding who was right between Copenhagenists and anti-Copenhagenists. According to Kaiser, at least in the US who was to become the new world center of science after the migration of many European scientists after the 1933, the new geo-political situation enhanced by the Second World War and the beginning of the nuclear age put a considerable pressure on physics education and training, with the side effect of virtually annihilating the area of the discussion on the philosophical implications of quantum mechanics, and physics in general. A new, strongly pragmatic attitude took over, exemplified by the positions of giants of the physics community such as Enrico Fermi or Richard Feynman, who advised implicitly or explicitly to leave philosophy aside:

> As for Feynman, he admonished his students at the California Institute of Technology that interpretive issues were all "in the nature of philosophical questions [and] not necessary for the further development of physics". These large-enrolment classes were by no means poorly executed. On the contrary, the lecture notes reveal the clarity and attention to detail that earned these leading physicists their well-deserved reputations. Yet for every additional example that Fermi, Bethe, Feynman and the others marched through at the blackboard – how to approximate the effects of an electric field on the energy levels of a hydrogen atom, or how to calculate the likelihood that two particles would scatter – they spent correspondingly less time encouraging their students to think hard about what all those fancy equations really said about the world. (Kaiser 2007, p. 32).[6]

Be as it may, a major factor for a new start in the interest for foundational issues in quantum mechanics, in a scientific community in which Bell was to become soon a leading figure, has been the establishment of Josef M. Jauch as a leader in the re-organization of the Swiss physics at the end of the 1950's. After completing his Diplom thesis under the supervision of Wolfgang Pauli in 1938, Jauch had a long research experience in the US and eventually, in 1959, received an offer for a chair in theoretical physics from the University of Geneva. As Charles P. Enz and Jagdish Mehra recall in the introduction to a volume for the Jauch sixtieth birthday,

> in Geneva, Josef Jauch was charged with reconstruction of the school of physics and the department of theoretical physics. Within four years he had brought the key people together and, eventually, resigned as the head of the physics institute in order to devote full attention to theoretical physics. *It was in Geneva that Jauch developed that other important research activity devoted to the foundations of quantum mechanics.*" (Enz, Mehra 1974, p. xv, my emphasis).

The interest for the foundations of quantum mechanics was united in the Jauch perspective to a rigorous mathematical attitude, and the outcome of these two strands was definitely his 1968 book, *The Foundations of Quantum Mechanics*. In the preface Jauch significantly expressed views that definitely turn out to resonate with the Kaiser characterization of the post-war training in physics vis-a-vis with issues on the very meaning of quantum theory:

> The pragmatic tendency of modern research has often obscured the difference between *knowing the usage of a language* and *understanding the meaning of its concepts*. There are many students everywhere who pass their examinations in quantum mechanics with top grades without really understand what it all means. Often it is even worse than that. Instead of learning quantum mechanics in parrot-like fashion, they may learn in this fashion only particular approximation techniques (such as perturbation theory, Feynman diagrams or dispersion relations), which lead them to believe that these useful techniques are identical with the conceptual basis of the theory. This tendency appears in scores of textbooks and in encouraged by some prominent physicists. This text, on the contrary, is not concerned with applications or approximations, but with conceptual foundations of quantum mechanics. (Jauch 1968, p. v, emphasis in the original).

---

[6] Kaiser support his general conclusions with a detailed study on the evolution of textbooks in US physics education between the late 1950's and the late 1970's (Kaiser 2007, pp. 32-33). In his 1991 book *Quantum Profiles*, Jeremy Bernstein also reports a personal investigation on the role of foundational issues in quantum mechanics textbooks: "In the early 1980s, before the impact of the work initiated by Bell had found its way into textbooks, I made a survey of seventeen standard textbooks on the quantum theory. I discovered that only one of them, David Bohm's *Quantum Theory*, published in 1951, made any reference at all to the paper of Einstein, Podolsky, and Rosen" (Bernstein 1991, p. 49).

At the beginning of 1963 Jauch visited CERN to give a seminar on a weakening of the no-hidden-variable result of von Neumann in 1932, a result that would have been published in a paper written with his collaborator Constantin Piron (Jauch, Piron 1963). Jauch and Bell had intense and fruitful discussions, the outcome of which would have appeared in the Bell assessment of the Jauch-Piron proposal in the section 4 of the first hidden variable paper[7]. Let us turn then to the Jauch-Piron results, not only to appreciate their scope and also what appeared to Bell its fatal drawbacks, but most of all to see them in connection with the locality-to-local-realism transition.

The Jauch and Piron approach[8] is in fact an instance of what in retrospect would have been called the *logico-algebraic*, or *quantum-logical* approach to quantum mechanics, firstly proposed in the von Neumann book of 1932 and in his 1936 paper with the American mathematician Garrett Birkhoff, and then developed in the late 1950's by the American mathematician George W. Mackey[9]. I will first outline the Jauch and Piron strategy in general terms, and then I will focus on those details that turn out to be especially relevant by a foundational point of view.

Jauch and Piron explicitly present their work as a follow-up to the von Neumann 1932 result: far from taking this result to be trivial or even circular[10], they wish to arrive at the very same conclusion on the basis of a weaker set of assumptions[11]. The Jauch and Piron approach proceeds according to the following pattern:

---

[7] As a piece of historical record, I recall the testimony of Nicolas Gisin, former student of Piron, that we find in the Freire Jr. book *Quantum Dissidents*: "As a matter of fact, Piron got upset with the success Bell obtained with these papers [the 1964 and 1966 Bell papers] …Bell's theorem was a taboo subject in Piron's circle." (2015, p. 299).

[8] At the beginning of the paper, Jauch and Piron explicitly justify their work by referring to "a renewed interest in a critique of the foundations of quantum mechanics" (Jauch, Piron 1963, p. 828), an expression which echoes the re-emergence of foundational interest in connection with the Jauch appointment in 1959 we mentioned above.

[9] von Neumann 1932, Birkhoff, von Neumann 1936, Mackey 1957. In the last footnote of their paper, Jauch and Piron mention explicitly the inspiring discussions and correspondence they had with Mackey. In his influential book *The Mathematical Foundations of Quantum Mechanics*, that would have been published just in that same year, Mackey clarified that "the aim of the book is to explain, or at least to illuminate, the essential aspects of classical and quantum mechanics from a point of view more congenial to pure mathematicians than that encountered in physics texts. In a rather unusual way, all physical concepts used are defined in terms of pure mathematics rather than physical connections except, of course, basic notions concerning space and time". (Mackey1963).

[10] They mention with disapproval (Jauch-Piron 1963, p. 828) the triviality charge and the circularity charge advanced respectively by De Broglie and Bohm in their 1957 books (De Broglie 1957, Bohm 1957).

[11] In the present analysis of the Bell 1966 I do not address the issue of the Bell assessment concerning the status of the von Neumann theorem. This is motivated not only by the existence of recent, excellent contribution on this specific issue (such as Acuña 2021), but also by the comparatively greater relevance of the Jauch-Piron theorem presuppositions to the issues of 'realism' (interpreted as a assumption on pre-existing values) and 'local realism'.

- the notion of *proposition for a physical system* is introduced, where a proposition is essentially a yes-no experiment, i.e. an experiment performable on the system that can produce one out of only two possible results;
- an algebraic structure is attributed to the set $\mathcal{L}$ of all relevant propositions, with special attention to a partial ordering relation holding for the propositions;
- the notion of *state* is defined as a suitable probability measure on $\mathcal{L}$, satisfying a number of conditions: the notion of *dispersion* is defined with respect to this notion of state, such that it is possible to introduce *dispersion-free* states, i.e. states that admit only 1 or 0 as probability values;
- finally, admitting hidden variables for the structure $\mathcal{L}$ is assumed to correspond to the existence of dispersion-free states for all the elements of $\mathcal{L}$.

The starting point is then the notion of proposition, that Jauch and Piron introduce in a way that was to become customary in the quantum logical tradition:

> [The] yes-no experiments […] are observables which can assume only one of two alternatives which we may designate by 1 or 0, yes or no, true or false. It is easy to exhibit a large number of examples of such observables and it is equally easy to show that the measurement of any measurable physical quantity can be reduced to the determination of a series of yes-no experiments. We shall in the following refer to such yes-no experiments as *propositions* of a physical system. (Jauch, Piron 1963, 829, emphasis in the original).

If $\mathcal{L}$ denotes the set of all meaningful propositions of a physical system, $\mathcal{L}$ is assumed to have an *algebraic structure*, firstly via the definition of an order relation $\subseteq$ such that, for $a, b \in \mathcal{L}$, $a \subseteq b$ if, whenever $a$ is true, also $b$ is true (Jauch, Piron 1963, p. 829). This relation is a *partial* order for $\mathcal{L}$, namely it is reflexive (for any $a \in \mathcal{L}$, $a \subseteq a$), antisymmetric (for any pair $a, b \in \mathcal{L}$, if $a \subseteq b$ and $b \subseteq a$, then $a = b$) and transitive (for any triple $a, b, c \in \mathcal{L}$, if $a \subseteq b$ and $b \subseteq c$, then $a \subseteq c$). In addition to the partial order, $\mathcal{L}$ is assumed to satisfy the following conditions:

- is a lattice, namely it is such that, for any for any pair $a, b \in \mathcal{L}$, the greatest lower bound $a \cup b$ and the least upper bound $a \cap b$ always exist and belong to $\mathcal{L}$;
- under the assumption that the operation $\cup$ and $\cap$ are extended to any subset $\{a_i\}$ of propositions to form the propositions $\cup_i a_i$ and $\cap_i a_i$, $\mathcal{L}$ contains the *absurd* proposition $\varnothing$ and the trivial proposition $I$, defined as

$$I = \cup_i a_i \ (a_i \in \mathcal{L}), \quad \varnothing = \cap_i a_i \ (a_i \in \mathcal{L})$$

- $\mathcal{L}$ is endowed with a *negation* operation $'$, such that for a given $a \in \mathcal{L}$, $a'$ is false whenever $a$ is true; moreover the negation operation satisfies for given $a, b \in \mathcal{L}$

$$(a')' = a;$$

$$a \cup a' = I, \quad a \cap a' = \varnothing;$$
$$(a \cup b)' = a' \cap b'$$

A set $\mathcal{L}$ satisfying the above conditions is called an orthocomplemented lattice. So far, these conditions can be satisfied by sets of propositions concerning both classical and quantum systems: the condition on which classical and quantum structures part their way is distributivity, namely the relations

$$a \cap (b \cup c) = (a \cap b) \cup (a \cap c)$$
$$a \cup (b \cap c) = (a \cup b) \cap (a \cup c)$$
(D)

hold in the classical domain but *not* in the quantum domain, due to the existence of pairs of propositions corresponding to mutually disturbing measurements (hence, in the Hilbert space machinery, to non-commuting projection operators). The replacement introduced by Jauch and Piron for the quantum case needs first of all the notion of *compatibility*: two propositions $a, b \in \mathcal{L}$ are said to be *compatible* (we will write $a \leftrightarrow b$) when they satisfy the relation

$$(a \cap b') \cup b = (b \cap a') \cup a.$$

It can be proved that a lattice $\mathcal{L}$ satisfies distributivity if and only if $a \leftrightarrow b$ for every pair of propositions $a, b \in \mathcal{L}$ : this illustrates even more clearly why a lattice of propositions concerning a quantum system cannot be distributive, since there certainly exist pairs of propositions – representing non-commuting measurements – that are not compatible. As a consequence, Jauch and Piron propose to weaken distributivity by replacing it with the following condition: for any pair $a, b \in \mathcal{L}$,

$$\text{if } a \subseteq b, \text{ then } a \leftrightarrow b. \tag{P}$$

The above structure, an orthocomplemented lattice satisfying (P), is called by Jauch and Piron a generalized *proposition system* (Jauch, Piron 1963, p. 831). In this structure, states $w$ are introduced as (generalized) probability measures $w:\mathcal{L}\to[0,1]$ on the lattice of propositions[12], satisfying the following conditions: for any $a, b \in \mathcal{L}$,

---

[12] As a matter of fact, states are presupposed when the very notion of proposition – along with its truth value *true* and *false* – is introduced. A proposition in the Jauch-Piron quantum-logical sense is the abstract analogue of an orthogonal projector **P** onto a linear subspace $P$ of the Hilbert space associated to the quantum system under scrutiny, namely the operator associated to an observable for the system *in the state* $\psi$ which can take only the values 1 – when $\psi \in P$ – and 0 – when $\psi \in P^\perp$, where $P^\perp$ is the subspace which is orthogonal to $P$ (see for instance David 2015, ch. 4). This is relevant also to negation since, for any $a \in \mathcal{L}$, $a'$ is the proposition such that $w(a') = 1 - w(a)$. When introducing the relation $a \subseteq b$ , Jauch and Piron claim that such relation "is the empirical analogue of the logical implication in the propositional calculus of ordinary logic" (p. 829). This is a curious claim, since the truth of a (material) implication in the propositional calculus of ordinary logic need

$$0 \leq w(a) \leq 1 \tag{1}$$

$$w(I) = 1, \ w(\emptyset) = 0 \tag{2}$$

$$\text{if } a \leftrightarrow b, \text{ then } w(a) + w(b) = w(a \cap b) + w(a \cup b) \tag{3}$$

$$\text{if } w(a_i) = 1 \text{ for any } i, \text{ then } w(\cap_i a_i) = 1, \text{ for any subset } \{a_i\} \text{ of propositions of } \mathcal{L} \tag{4}$$

$$\text{if } a \neq I, \text{ then there is a state } w \text{ such that } w(a) \neq 0. \tag{5}$$

If $w_1$ and $w_2$ are two different states then, given a proposition $a \in \mathcal{L}$, the state $w$

$$w(a) = \lambda_1 w_1(a) + \lambda_2 w_2(a), \qquad \text{with } \lambda_1, \lambda_2 > 0, \lambda_1 + \lambda_2 = 1$$

is said to be a *mixture*. Given a state $w$ and a proposition $a \in \mathcal{L}$, the quantity

$$\sigma_w(a) = w(a) - w^2(a)$$

is defined as the *dispersion* of $w$ on the proposition $a$; then, a state $w$ is called *dispersion-free* when, for any $a \in \mathcal{L}$, $\sigma_w(a) = 0$. Finally, a system $\mathcal{L}$ is said to admit hidden variables when *every* state definable on $\mathcal{L}$ is a mixture of dispersion-free states. Essentially under these assumptions, Jauch and Piron prove that the hypothesis that a proposition system $\mathcal{L}$ admits hidden variables implies the mutual compatibility ($a \leftrightarrow b$) of *every* pair of propositions $a$, $b \in \mathcal{L}$: but if a proposition system $\mathcal{L}$ is a *quantum* system, then there certainly exist pairs of *incompatible* propositions, a circumstance from which the impossibility of hidden variables for $\mathcal{L}$ follows (Jauch, Piron 1963, p. 835, corollary 3)[13].

A crucial point in the above result, a point on which the critical review in Bell 1966 was especially focused, concerns the condition (4), namely if $w(a_i) = 1$ for any $i$, then $w(\cap_i a_i) = 1$, for any subset $\{a_i\}$ of propositions of $\mathcal{L}$.

Relativized to a pair of $a, b \in \mathcal{L}$, this condition requires that whenever $w(a) = 1$ and $w(b) = 1$, also $w(a \cap b) = 1$ holds; in other words, when a measurement of the propositions provides an answer yes with certainty for each $a$ and $b$ individually, also a measurement of $a$ AND $b$ will provide an answer yes with certainty. When $a$ and $b$ are compatible, this is a consequence of the condition (3). For suppose that $w(a) = 1$ and $w(b) = 1$; then, jointly with (3), we have

---

not be restricted to the situation in which, whenever the antecedent is true, the consequent is also true (in order for an implication $a \rightarrow b$ to be true, it is sufficient that $a$ is false or $b$ is true).

[13] The same result was derived in 1968 by Gudder under still weaker assumptions: the proposition system $\mathcal{L}$ was taken by Gudder to be not necessarily a lattice, but only an orthocomplemented partially ordered set which is complete with respect to compatible elements: "The importance of this generalization physically lies in the fact that there seems to be no experimental evidence that [$\mathcal{L}$] should be complete with respect to arbitrary elements, although completeness with respect to compatible elements is physically much more reasonable" (Gudder 1968, p. 319).

$$w(a) + w(b) = 2 = w(a \cap b) + w(a \cup b),$$

from which $w(a \cap b) = 1$, since the value of each $w(a \cap b)$ and $w(a \cup b)$ can equal 1 at most. But, as Jauch and Piron emphasize, the condition (4) does *not* follow from (3) when $a$ and $b$ fail to be compatible, which is the reason why (4) is in fact *postulated*: in this version, the condition is referred to as (4)°.

Leaving aside the original Bell objection, according to which the validity of the condition (4) in standard Hilbert space QM is a *contingent* circumstance that need not hold for *any* completion of QM (an objection similar in spirit to the objection raised to the von Neumann 1932 no-hidden-variable result, Bell 1987, pp. 5-6), I argue that we can single out a Pre-Existing-Values kind of assumption underlying the Jauch-Piron condition (4). If, in line with a Pre-Existing-Values assumption, we interpret $w(a) = 1$ and $w(b) = 1$ as *grounded* in the fact that those values for $a$ and $b$ *were already there*, then it is fairly obvious that also the conjunction of pre-existing values will give rise to probability assignment equal to 1. In this perspective, it turns out to be rather natural to assume that the validity of both $w(a) = 1$ and $w(b) = 1$ justifies the validity of $w(a \cap b) = 1$ and, in this sense, the naturalness with which Jauch and Piron assume condition (4) would be due to unconsciously taking for granted that a hidden variable theory *must* be a Pre-Existing-Values satisfying theory. This is consistent also with the above condition according to which a system $\mathfrak{L}$ is said to admit hidden variables when every state definable on $\mathfrak{L}$ is a mixture of dispersion-free states: as Jauch and Piron state explicitly in a different paper, a system $\mathfrak{L}$ is said to admit hidden variables when every state definable on $\mathfrak{L}$ is a mixture of dispersion-free states *for all physical quantities* (Jauch, Piron 1968, p. 228): the most natural justification for this definition is exactly in terms of pre-existing values, since it is this pre-existence that once again grounds the lack of dispersion for *all* physical quantities.

But more can be said in support of this claim if we take into account a criticism leveled against the Jauch-Piron approach by David Bohm and Jeffrey Bub in their 1966 paper (Bohm, Bub 1966)[14]. They remark that, when $a$ and $b$ are two *incompatible* propositions, we have

$$a \cap b = a' \cap b = a \cap b' = a' \cap b' = \varnothing$$

Therefore, when $a$ and $b$ are incompatible, there can be no *quantum* state $w$ that assigns probability 1 to the measurement of $a$ and $b$ separately and *also* probability 1 to $a \cap b$ (Bohm, Bub 1966, p. 474)! To require the condition (4)° *anyway* – as Jauch and Piron do – amounts then to the claim that, although there is no *quantum* state that assigns certainty to the conjunction of $a \cap b$ when the measurement results on $a$ and $b$ separately are certain, nevertheless the conjunction $a \cap b$ has certainly a yes-result because it is a conjunction of the *pre-existing values* for $a$ and $b$. Bohm and Bub themselves implicitly hint at this, when they emphasize how contrived the requirement of (4)° appears to them, in spite of the above unfortunate mathematical implication: "[…] the very essence of a hidden variable theory is that, for a completely specified state of the system, *in which the values of the hidden variables are all determined*, the result of the measurement of any observable can be predicted with certainty." (Bohm, Bub 1966, p. 474, my emphasis [15]).

---

[14] Bohm and Bub provide in their paper several different criticisms of the Jauch-Piron approach, but I consider here only the one which is relevant to my perspective, focused on the assumption of Pre-Existing Values.

## 3      'Local realism' *in* the Bell theorem (1964)

If the first Bell paper had a weird editorial fate and gained popularity with a year of publication which comes *three* years after its actual production, also the second, 1964 paper lived a peculiar editorial story. We know from the events connected with the first paper that Bell was in the United States at that time. After completing the manuscript, he had to choose to which journal to submit it but he lacked resources for paying the page charges that renowned physics journals requested from authors (he was a guest of colleagues in the US and he did not want to ask for additional fundings, even less for a non conventional paper such as that). This lack of resources led him toward a journal that not only did not impose page charges but also paid the authors for publication. It was *Physics Physique физика*, an inter-disciplinary journal published during the period 1964-1968 and founded by P. W. Anderson (Nobel Prize in 1977) and B. T. Matthias, whose attempt was to "alleviate the increasing problems of the communication of science", as they claimed in their Editorial Foreword. The journal was short-lived but published papers by eminent physicists, such as Gell-Mann, Ne'eman and of course Bell, whose manuscript was read and accepted for publication by Anderson himself, ironically also because he took it as a refutation of Bohm's hidden variable theory[16].

There is a clear sense in which, both historically and conceptually, the first Bell paper paved the way to the second, which contains the first instance of what would have become *the* Bell theorem (Bell 1964). That is, the first Bell paper shows that the insistence on pre-existence of values does nothing but obscure the heart of the matter: a condition of *locality*, that an already existing formulation of hidden variable theory  - the Bohm theory – fails to satisfy and whose *general* status for *any* possible theory incorporating quantum predictions was still to be assessed. In the first page of his paper, Bell summarizes the EPR-Bohm incompleteness argument in order to state unambiguously the premises from which his own non-locality theorem is to proceed. I will start first from the informal wording that Bell himself employs in stating the aim of his article, and I will proceed to a coincise, step-by-step formulation of the EPR-Bohm argument in order to show that a condition of pre-existence of values is *derived* and not *assumed*. Finally I will quote the Bell summary of the situation, a summary that states clearly the derivative character of a pre-existence of values.

---

[15] It is curious to remark that in the 1968 paper by Jauch and Piron mentioned above (Jauch, Piron 1968), which is supposed to be a direct and explicit reply to the Bohm-Bub criticisms, Jauch and Piron do not address this rather decisive point, nor provide in fact a detailed discussion of any of the points raised by Bohm and Bub. The latter provided a counter-reply in Bohm, Bub 1968.

[16] That this was the Anderson view of the Bell result comes from a personal communication from Anderson to David Wick (Wick 1995, p. 90, footnote 103).

John S. Bell opens his article as follows:

> The paradox of Einstein, Podolsky and Rosen was advanced as an argument that quantum mechanics could not be a complete theory but should be supplemented by additional variables. These additional variables were to restore causality and locality. In this note that idea will be formulated mathematically and shown to be incompatible with the statistical predictions of quantum mechanics. *It is the requirement of locality, or more precisely that the result of a measurement on one system be unaffected by operations on a distant system with which it has interacted in the past, that creates the essential difficulty.* (Bell 1964, in Bell 1987, p. 14, my emphasis)

Let us consider the point in a specific formulation of the EPR argument in the Bohm version, in which we have a composite quantum system $S_1+S_2$ of a pair of spin-1/2 particles $S_1$ and $S_2$. The composite system is prepared at a time $t_0$ in the singlet state $\Psi$

$$\Psi = 1/\sqrt{2}\ (|1,+>_n |2,->_n - |1,->_n |2,+>_n),$$

where **n** denotes a generic spatial direction. We take into account the measurements concerning the spin components along given directions, whose possible outcomes are only two (conventionally denoted by '+1' and '-1'). We assume also that the spin measurements on $S_1$ and $S_2$ are performed when $S_1$ and $S_2$ occupy two mutually isolated spacetime regions $R_1$ and $R_2$. According to QM, we know that if the state of $S_1+S_2$ at time $t_0$ is $\Psi$, then the (reduced) states of the subsystems $S_1$ and $S_2$ at time $t_0$ are respectively

$$\rho(1,\Psi)=1/2(\mathbf{P}_{|1,+>\,n} + \mathbf{P}_{|1,->\,n}), \quad\quad\quad \text{(RS)}$$
$$\rho(2,\Psi)=1/2(\mathbf{P}_{|2,+>\,n} + \mathbf{P}_{|2,->\,n}),$$

('RS' stands for 'reduced states') so that, for any **n,**

$$\text{Prob}_{\rho(1,\Psi)}\ (\text{spin}\ \mathbf{n}\ \text{of}\ S_1 = +1) = \text{Prob}_{\rho(1,\Psi)}\ (\text{spin}\ \mathbf{n}\ \text{of}\ S_1 = -1) = 1/2$$
$$\text{Prob}_{\rho(2,\Psi)}\ (\text{spin}\ \mathbf{n}\ \text{of}\ S_2 = +1) = \text{Prob}_{\rho(2,\Psi)}\ (\text{spin}\ \mathbf{n}\ \text{of}\ S_2 = -1) = 1/2$$

Moreover, if we perform at a time $t$ a spin measurument on $S_1$ along **n** with outcome +1 (–1), a spin measurement on $S_2$ along **n** at a time $t' > t$ will give with certainty the outcome –1 (+1), namely for any **n** ('**AC**' stands for '**A**nti**C**orrelation')

$$\text{Prob}_\Psi\ [(\text{spin}\ \mathbf{n}\ \text{of}\ S_1 = +1)\ \&\ (\text{spin}\ \mathbf{n}\ \text{of}\ S_2 = -1)] = \quad\quad\quad \text{(AC)}$$

$$\text{Prob}_\Psi [(\text{spin}_\mathbf{n} \text{ of } S_1 = -1) \,\&\, (\text{spin}_\mathbf{n} \text{ of } S_2 = +1)] = 1.$$

Let us suppose now to perform at time $t_1 > t_0$ a spin measurement on $S_1$ with outcome +1. Therefore, according to (**AC**), a spin measurement on $S_2$ along **n** at a time $t_2 > t_1$ will give with certainty the outcome – 1. Let us suppose now to assume the following condition:

> **REALITY** – If, without interacting with a physical system *S*, we can predict with certainty - or with probability 1 - the value **q** of a quantity **Q** pertaining to *S*, then **q** represents an objective property of *S* (denoted by [**q**]).

Then, for $t_2 > t_1$ [**spin**$_\mathbf{n}$ = –1] represents an objective property of $S_2$. But might the objective property [**spin**$_\mathbf{n}$ = –1] of $S_2$ have been somehow "created" by the spin measurement on the distant system $S_1$? The answer is negative if we assume the following condition:

> **LOCALITY** – No objective property of a physical system *S* can be influenced by operations performed on physical systems that are isolated from *S*.

At this point, **LOCALITY** allows us to state the existence of the objective property [**spin**$_\mathbf{n}$ = –1] for the system $S_2$ also at a time $t'$ such that $t_0 > t' > t_1$. Namely, if we assume that the measurement could not influence the validity of that property at that time, it follows that the property *was holding already at time $t'$*, a time that *precedes* the measurement performed on the other subsystem. But at time $t'$ the state of $S_1+S_2$ is the singlet state $\Psi$, therefore according to (**RS**) the state of $S_2$ is the reduced state $\rho(2,\Psi)=1/2(\mathbf{P}_{|2,+> \mathbf{n}} + \mathbf{P}_{|2,-> \mathbf{n}})$, that prescribes for the property [**spin**$_\mathbf{n}$ = –1] of $S_2$ only a probability 1/2. Let us consider finally the following condition:

> **COMPLETENESS** – Any objective property of a physical system *S* must be represented within the physical theory that is supposed to describe *S*.

It follows that there exist properties of physical systems that, according to the **REALITY** condition are objective – like [**spin**$_\mathbf{n}$ = –1] for $S_2$ – but that QM does not represent as such: therefore QM is not complete.

As should be clear from the above presentation, the counterfactual reasoning here concerns the question: "Might the objective [**spin**$_\mathbf{n}$ = –1] of $S_2$ have been somehow "created" by the spin measurement on the distant system $S_1$?". But *given* **LOCALITY**, the only acceptable possibility for justifying the existence of the property [**spin**$_\mathbf{n}$ = –1] for the system $S_2$ also at a time $t'$ is its pre-measurement definiteness, but such definiteness is *entailed* by **LOCALITY** and nowhere the argument needs to *assume independently* that spin properties are definite. This is equivalent to stress that assuming **REALITY** is not equivalent to assuming at the outset the existence of definite, pre-measurement properties: **REALITY** is simply a sufficient criterion for a property of physical system to be objective, namely holding in a measurement-independent way.

And here is the Bell summary:

> Consider a pair of spin one-half particles created somehow in the singlet spin state and moving freely in opposite directions. Measurements can be made, say by Stern-Gerlach magnets, on selected components of the spins $\sigma 1$ and $\sigma 2$. If measurement of the component $\sigma 1 \cdot \mathbf{a}$, where $\mathbf{a}$ is some unit vector, yields the value +1 then, according to quantum mechanics, measurement of $\sigma 2 \cdot \mathbf{a}$ must yield the value −1 and vice versa. Now we make the hypothesis, and it seems one at least worth considering, that if the two measurements are made at places remote from one another the orientation of one magnet does not influence the result obtained with the other. Since we can predict in advance the result of measuring any chosen component of $\sigma 2$, by previously measuring the same component of $\sigma 1$, *it follows* that the result of any such measurement must actually be predetermined. Since the initial quantum mechanical wave function does not determine the result of an individual measurement, this predetermination implies the possibility of a more complete specification of the state. (Bell 1964, in Bell 1987, pp. 14–15, my emphasis)

As should be clear from a fair reading of the Bell original article, the Bell theorem starts exactly from the alternative established by the EPR-Bohm argument—namely, locality and completeness cannot stand together—and goes for the proof that, *whatever form the completability of quantum mechanics might assume*, the resulting theory cannot preserve the statistical predictions of quantum mechanics and be local at the same time: this means that neither a pre-existing-property assumption nor a determinism assumption are assumed in the derivation of the original Bell inequality. As Bell himself clearly stresses:

> It is important to note that to the limited degree to which *determinism* plays a role in the EPR argument, it is *not assumed* but *inferred*. What is held sacred is the principle of 'local causality'—or 'no action at a distance'. [. . .] It is remarkably difficult to get this point across, that determinism is not a *presupposition* of the analysis." [. . .] My own first paper on this subject [Bell refers here to his 1964 paper] starts with a summary of the EPR paper *from locality to deterministic hidden variables*. But the commentators have almost universally reported that it begins with deterministic hidden variables. (Bell 1982, in Bell 2004[2], pp. 143, 157, footnote 10, emphasis in the original)

But there is more to this question. That a pre-existence condition cannot be a reasonable *independent* assumption of any allegedly 'objective' theory of quantum phenomena had been already suggested in the Bell 1966 paper. In showing that all existing no-hidden variable theories proofs at the time required assumptions that it was not reasonable to require from any hypothetical completion of quantum theory, Bell argued:

> It will be urged that these analyses [i.e. the above mentioned proofs] leave the real question untouched. In fact it will be seen that these demonstrations require from the hypothetical dispersion free states, not only that appropriate ensembles thereof should have all measurable properties of quantum mechanical states, *but certain other properties as well*. These additional demands appear reasonable when results of measurement are loosely identified with properties of isolated systems. They are seen to be quite unreasonable when one remembers with Bohr 'the impossibility of any sharp distinction between the behaviour of atomic objects and the interaction with the measuring instruments which serve to define the conditions under which the phenomena appear'. (Bell 1966, in Bell 2004[2],, pp. 1–2, my emphasis)

If the pre-existence of values were an *independent* assumption of any hidden variable theory, Gleason-Bell-Kochen & Specker would have already proved their incompatibility with quantum mechanics *needless of any locality requirement*. But, as Bell showed, there is little significance in testing against quantum theory a theory (be it local or non-local) that is supposed to satisfy a condition that we already know quantum mechanics cannot possibly and reasonably satisfy.

In addition to the logical status of the premises of the Bell theorem, there was a deep motivation for restricting the attention to locality. In the area of investigations opened nearly half a century ago by John S. Bell, the question naturally arose of what would have been the implications of *extending* quantum mechanics, in view of the emergence of phenomena that were not easy to accommodate within a familiar view of the physical world, non-locality being the most urgent case. Due to the unavoidable existence of entangled states – something that makes quantum mechanics a non-local theory in a fundamental sense due to the linearity of the theory, of which entanglement is a consequence – it has seemed plausible to put things in the following way: let us ask whether quantum mechanics might be seen as a 'fragment' of a more general theory which – at a 'higher' level – may recover that locality that turns out not to hold at the strictly quantum level. One of the *strong points* of the original Bell strategy that led to the Bell-named theorem was exactly that this hypothetical extension was confined to the locality/non-locality issue and needed not say *anything* on further details concerning 'realistic' or 'non-realistic' properties, states or whatever: in addition to being useful for the economy of the theorem, this point was absolutely plausible since it makes sense to require from the extension the only condition that we are interested to add to the new hypothetical super-theory, namely locality.

## 4 'Local realism' *after* the Bell theorem: the 1969 Clauser-Horne-Shimony-Holt paper

The next relevant step in the chain, leading from the first Bell paper on the subject (the 1966 paper) toward the experimental confirmation of the divergence between quantum mechanics and more and more general versions of a local theory able to incorporate the quantum predictions, is the paper published by John F. Clauser, Michael A. Horne, Abner Shimony and Richard A. Holt in 1969, that contains the first generalized inequality known as CHSH inequality[17]. The paper opens with a reconstruction of the situation at the point where the EPR argument had left it: this reconstruction, in fact, shows how a misguided reading of the EPR argument as to the status of its main assumptions starts to orient the debate toward a 'local-realistic' view. Therefore this paper not only is – according to what John Clauser claims in retrospect (Clauser 2002, p. 80, fn 14; I will return later to this paper) – the *locus* in which the expression *Bell's theorem* appears for the first time; it appears to be also the first paper that, although implicitly, starts to give credit to the idea that what the Bell theorem is about is the *conjunction* of locality *and* realism, where realism is once again the assumption of pre-existing values for the relevant observables. To see why, let us quote in full the opening lines of the CHSH 1969 paper:

> Einstein, Podolsky, and Rosen (EPR) in a classic paper presented a paradox which led them to infer that quantum mechanics is not a complete theory. They concluded that the quantum mechanical description of a physical system should be supplemented by postulating the existence of 'hidden variables', the specification of which would predetermine the result of measuring any observable of the system. They believed the predictions of quantum mechanics to be correct, but only as consequences of statistical distributions of the hidden variables.

If we consider the concluding lines of the original EPR paper, however, which reads

> While we have shown that the wave function does not provide a complete description of physical reality, we left open the question of whether or not such a description exists. We believe, however, that such a theory is possible. (EPR, p. 780) we realize that CHSH seriously over-interpret in purely textual terms the EPR conclusion, since they assume the equivalence

| CHSH – "the quantum mechanical description of a physical system should be supplemented by postulating the existence of 'hidden variables', the specification of which would predetermine the result of measuring any observable of the system." | 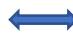 | EPR – "We believe, however, that [a complete description of physical reality] is possible." |
|---|---|---|

In other words, CHSH take the *generic hope* of finding an alternative theoretical framework, expressed by the EPR wording, to be strictly equivalent to a requirement concerning the *existence* of hidden variables – namely, in our terms, the requirement of pre-existing values for all relevant observables – that would *pre-determine* any measurement result, an equivalence that seems far from established, at a minimum.

---

[17] A lively historical reconstruction of the events that led the four authors to join their efforts for the production of the paper is contained in the Chapter 3 of Kaiser 2011.

It is interesting to remark that the extent to which the EPR plea for a 'complete' theory should be read automatically in terms of a hidden variable theory of the sort envisaged in the CHSH opening lines is addressed again in the Appendix of a Bell 1976 paper, entitled "Einstein-Podolsky-Rosen Experiments" (Bell 1976, in Bell 2004[2], pp. 81-92): this seems the only place in which *Bell himself* gives credit to this reading. The Appendix, in fact, is the Bell summary of a controversy between himself and the well-known and influential historian of physics Max Jammer, who had expressed divergent views in his 1974 book *The Philosophy of Quantum Mechanics*. In the chapter 7 of the book, devoted to hidden variable theories, Jammer presents his view of the relation between the Einsteinian reservations on quantum mechanics and the hidden variable program, wondering to what extent the former might have been the main factor for the development of the latter:

> Although the Einstein-Podolsky-Rosen incompleteness argument was undoubtedly one of the major incentives for the modern development of hidden variable theories, it would be misleading to regard Einstein, as some authors do, as a proponent or even as "the most profound advocate of hidden variables "[18] True, Einstein was sympathetically inclined toward any efforts to explore alternatives, and as such the ideas of de Broglie and of Bohm, but he never endorsed any hidden variable theory. […] No doubt, Einstein's criticisms, and in particular his work with Podolsky and Rosen greatly contributed to the development of hidden variable theories, just as Mach's ideas contributed to the rise of Einstein relativity: but, as is not uncommon in the history of physics, the intellectual originator of a theory does not necessarily identifies himself with its full-fledged development. (Jammer 1974, pp. 254-255)

In particular, Jammer points at the Bell 1964 paper as a major source for what he takes to be a misleading view, when in a footnote to the above quoted passage he refers to the opening lines of the Bell paper (already quoted in section 3), where the author writes that "[t]he paradox [i.e. the EPR argument] was advanced *as an argument that quantum mechanics should be supplemented by additional variables*." (Jammer 1974, p. 254, fn 3, my emphasis).[19] In the Appendix to his 1976 paper Bell, after quoting *with approval* the very same Shimony characterization of Einstein as "the most profound advocate of hidden variables" that Jammer had attacked, replies to the Jammer footnote. Still in this footnote, Jammer had further claimed that some Einstein remarks, taken by Bell 1964 to support his view of hidden variable theories as *direct* consequence of the EPR argument, could not be read as "confessions of the belief in the necessity of hidden variables" (Jammer 1974, p. 254, fn 3). What was the Einstein remark?

---

[18] The reference is to Shimony 1971 (reprinted in Shimony 1993, p. 87).
[19] In the same footnote Jammer writes that the way in which CHSH present the EPR argument in the opening lines of the 1969 paper is "under Bell's influence".

> But on one supposition we should, in my opinion, absolutely hold fast: the real factual situation of the system $S_2$ is independent of what is done with the system $S_1$, which is spatially separated from the former. (Einstein 1959, p. 85)

In order to justify his use of this remark at the beginning of his 1964 paper, Bell writes that "the object of this quotation was to recall Einstein's deep commitment *to realism and locality, the axioms of the EPR paper*." (Bell 1976, in Bell 2004[2], p. 89, my emphasis), and it is important to recall that Bell quotes the Einstein remark in a footnote to a passage where he claims that "these additional variables were to restore to the theory *causality* and locality." (Bell 1964, in Bell 2004[2], p. 14, my emphasis), where 'causality' clearly, altough in a rather sloppy way, plays the role of what Bell calls 'realism' in the 1976 Appendix. As is clear from this wording, then, Bell himself cannot avoid to run afoul of the very same mistake that we analyze in the present work, and that he denounces himself most clearly in his 1982 paper "Bertlmann's socks and the nature of reality":

> It is important to note that to the limited degree to which *determinism* plays a role in the EPR argument, it is *not assumed* but *inferred*. What is held sacred is the principle of 'local causality'— or 'no action at a distance'. [. . .] It is remarkably difficult to get this point across, that determinism is not a *presupposition* of the analysis." (Bell 1982, in Bell 2004[2], p. 143, emphasis in the original).

And in a footnote few lines later he explicitly writes

> "My own first paper on this subject [Bell refers here to his 1964 paper] starts with a summary of the EPR paper *from locality to deterministic hidden variables*. But the commentators have almost universally reported that it begins with deterministic hidden variables." (Bell 1982, in Bell 2004[2], p. 157, emphasis in the original).

Remarkably, in saying in the 1976 Appendix that 'realism' is *also* an 'axiom' of the EPR paper, Bell is doing here exactly what "the commentators have almost universally reported"!

But let us return to the CHSH paper, whose explicit aim is to propose a more realistic version of the Bell 1964 framework, since the *strict* anti-correlation encoded into the spin singlet state turned out to be experimentally unfeasible. The CHSH framework considers an ensemble of correlated pairs of particles such that each member of the pair is supposed to reach an apparatus. In fact, the CHSH framework started to employ photons, that are supposed to be directed toward a linear polarization filter, associated with a given axis: photons with polarization parallel to the chosen axis will pass through the filter and will reach a detector, whereas the photons with polarization orthogonal to the chosen axis will be absorbed[20]. $I_a$ and $I_b$ denote the apparatuses, where *a* and *b* denote adjustable parameters (like the directions along which the axis of the polarizers are set), and in each apparatus the particle selects one of two channels. If +1 and – 1 denote the selection of the first or the second channel, respectively, CHSH represent with *A*(*a*) and *B*(*b*) the results of these selections. Since the members of each pair were joint and interacting at the source, we can safely assume that there is a state description that (i) encodes the information that the two particles typically shared at the source, and (ii) takes into account the contribution of a 'more complete' description allowed by hidden variables. The point of what this 'contribution' is supposed to mean is delicate, so that it is useful here to quote CHSH in full:

> Suppose now that a statistical correlation of *A*(*a*) and *B*(*b*) is due to information carried by and localized within each particle, and that at some time in the past the particles constituting one pair were in contact and communication regarding this information. The information, *which emphatically is not quantum mechanical*, is part of the content of a set of hidden variables, denoted collectively by λ. The results of the two selections are then to be *deterministic functions A*(*a*, λ) and *B*(*b*, λ). (CHSH 1969, p. 881, my emphasis)

Clearly, this formulation is supposed to do justice to the idea that the hidden variable theory should 'complete' quantum mechanics, to the effect that in principle the pair (*a*, λ) [(*b*, λ)] suffices to fix in each case whether *A*(*a*, λ) [*B*(*b*, λ)] equals +1 or – 1. Moreover CHSH assume the following conditions:

**LOCALITY**

$$A(a, b, \lambda) = A(a, \lambda)$$
$$B(b, a, \lambda) = B(b, \lambda)$$

**INDEPENDENCE**

The set λ has a normalized probability distribution ρ(λ) that is independent from the parameters *a* and *b*.

A so-called *correlation function P*(*a*, *b*) is then introduced such that, according to LOCALITY and INDEPENDENCE,

$$P(a, b) = \int_\Lambda A(a, \lambda)\, B(b, \lambda)\, \rho(\lambda)\, d\lambda$$

Finally, in order to substantiate the aim to provide a framework that might turn out to be more suitable to experimental test, CHSH assume less-than-strict correlation, namely that there are parameters *b* and *b'* such that $P(b, b') = 1 - \delta$, with $0 \leq \delta \leq 1$: "Experimentally interesting cases will have δ close to but not equal to zero. Here we avoid Bell's experimentally unrealistic restriction that for some pair of parameters $b$ and $b'$ there is perfect correlation (i.e., $\delta = 0$)." (CHSH 1969, p. 881). On the basis of the above assumptions, CHSH are then able to derive an inequality

$$|P(a, b) - P(a, c)| + |P(b, b') - P(b', c)| \leq 2$$

for which a suitably proposed experimental test will vindicate quantum mechanics against the envisaged local hidden variable theory.

---

[20] As is well known, at the beginning of the era of experimental tests of Bell-type inequality with photons, the detector efficiency was very low, that is less than 20% (Whitaker 1996, p. 261).

The CHSH paper, to whose opening lines I referred to above as an exercise of over-interpretation of the EPR paper conclusion, suggests a new, more general experimental framework, in which correlation is not strict as in the original 1964 Bell framework, in which the derived Bell inequality was *de facto* experimentally untestable:

> Bell's theorem has profound implications in that it points to a decisive experimental test of the entire family of local hidden-variable theories. The aim of this paper is to propose explicitly such an experiment. For this purpose, we first present a generalization of Bell's theorem which applies to realizable experiments. (CHSH 1969, p. 880-881).

In doing this, CHSH assume *both* that

(i) the EPR argument *postulates* the existence of hidden variables that predetermine any measurement outcome,

and that

(ii) the theorem in the 1964 Bell paper, inasumuch as it starts from where the EPR argument ends, shares with EPR the *postulation* of those very same hidden variables.

But, as we have emphasized in the preceding section, the 'postulate' of the existence of hidden variables as pre-existing values plays no independent role whatsoever either in the original EPR paper or in the derivation of the Bell inequality, since it is a *consequence* of locality, and it is in *this* sense that the Bell theorem starts from where the EPR argument ends. Moreover, even the argument that the relaxation of strict correlation 'requires' a determinism (in terms of 'pre-existence') assumption is unjustified: for it can be shown unequivocally that *also in the derivation of the CHSH inequality* the postulation of pre-existing values as an independent assumption plays no logical role, since the above assumptions of LOCALITY and INDEPENDENCE are jointly sufficient to derive that inequality[21].

---

[21] A neat derivation of the CHSH inequality, showing that "no assumptions of determinism or of a 'classical behavior' are being involved" can be found in Brunner *at al.* 2014, p. 421-422. In fact, due to the irrelevance of 'determinism' in the CHSH introduction of 'deterministic functions', the CHSH theoretical framework appears to be equivalent to the Bell 1971 framework, which is usually taken to be the first instance of a *stochastic* hidden variable theory. In CHSH the determinism would be there, but due to the limits in the detection, it cannot be really observed in action and, in any case, it plays no role in the derivation of the inequality. On the other hand, the Bell stochastic hidden variable theory assumes explicitly a *probabilistic* connection between $\lambda$ and the measurement values.

An extremely interesting source to take into account, in order to realize how controversial the road to 'local realism' turns out to be, is a recollection of John F. Clauser himself, contained in a paper appeared in a 2002 collective book devoted to the implications of Bell's results[22]. In a section entitled *Generalization of the Bell and CHSH Results to Constrain Local Realism and Space-Time*, and in response to a question that he formulates as "What are the fundamental assumptions underlying Bell's Theorem?" (Clauser 2002, p. 82), he claims:

> Following EPR's lead, both Bell and CHSH initially had assumed that determinism is the basic underlying assumption for Bell's Theorem. This notion follows from EPR's *definition* of an "element of reality", as per "If, without in any way disturbing a system, we can predict with certainty (i.e. with probability equal to unity) the value of a physical quantity, then there exists an element of physical reality corresponding to this physical quantity." (Clauser 2002, p. 83, emphasis in the original).

Firts of all then, as is clear from this passage, in Clauser's view the EPR (necessary) condition for elements of physical reality amounts not only to clarify what it means to be an element of physical reality but *also to assume the independent existence of such elements of reality*, something that does not follow from the very EPR definition (see again our reconstruction of the EPR argument in section 3). But immediately after Clauser writes:

> While predictions with certainty are possible for idealized systems, unfortunately, no such predictions can be made for realizable systems (except, of course, for death and taxes). Following EPR, Bell effectively derives determinism in his 1964 paper for the EPR particles, via a use of the at least one data point with a perfect correlation. Unfortunately and conversely, determinism cannot be derived via this reasoning for any real system, which, of course, has no such point with Bell's perfect correlation, and thus no predictions with probability equal to unity. Given this fact but still following EPR's ideas, CHSH simply and explicitly assumed determinism to hold for the purposes of their derivation. (Clauser 2002, p. 83).

That is, Clauser evokes the CHSH proposal of a more realistic experimental framework as compared to the Bell 1964 framework, emphasizing that this proposal *requires* to assume determinism and making explicit that the Bell 1964 *did not* require it, a statement that is plainly inconsistent with the above passage ("Following EPR's lead, both Bell and CHSH initially had assumed that determinism is the basic underlying assumption for Bell's Theorem."). Not only that. Immediately after, Clauser recalls that in working with Mike Horne on a well-known paper that would have appeared in 1974

---

[22] The status of this book, *Quantum (Un)Speakables. From Bell to Quantum Information*, is actually somewhat paradoxical, starting from the very title, since the vast majority of the authors (and the editors!) subscribe to interpretations and viewpoints on quantum mechanics that are very far removed from the Bell stance on these issues: the reference to quantum information speaks for itself.

Mike Horne and I noticed the fact that the CHSH prediction also appears to hold for models in which determinism is neither assumed, nor in fact, even holds. This fact suggested to us that an assumption of tha determinism is not really necessary for Bell's Theorem to apply." (Clauser 2002, p. 83), a statement that is again in tension with the above claim according to which "CHSH simply and explicitly assumed determinism to hold for the purposes of their derivation"!

## 5      'Local realism' *after* the Bell theorem: the 1974 Clauser-Horne paper and the 1978 Clauser-Shimony paper

The next step is undoubtedly represented by the paper that Clauser and Horne (CH from now on) publish in 1974, entitled "Experimental consequences of objective local theories" (Clauser, Horne 1974). It is worth examining the very opening of the paper, in order to realize once again how logically correct and incorrect perspectives get mixed and form sometimes an inextricable mixture.

> Two papers by Bell have shown that the statistical predictions of quantum mechanics, for certain spatially separated yet correlated two-particle systems, are incompatible with a broad class of local theories. Bell's earlier paper considers the consequences of a physically reasonable locality condition within the domain of an ideal *Gedankenexperiment*. He demonstrates that any theory which satisfies the locality condition must also be deterministic if certain quantum-mechanical predictions are valid for the idealized case. (Clauser, Horne 1974, p. 526)

The two Bell papers mentioned here by CH are the well-known 1964 Bell paper and a 1971 Bell paper in which Bell presents for the first time a form of *stochastic* hidden variable theory (Bell 1971),. The passage above refers to the first, the 1964 paper, and seems to confirm a correct reading of it, according to which it is the independent locality assumption that allows one to derive determinism: CH clearly endorse the validity of this implication when they aptly write that "any theory which satisfies the locality condition *must also be deterministic* if certain quantum-mechanical predictions are valid for the idealized case". Strangely enough, in the very subsequent lines they turn out to be much more ambiguous:

> Bell' s further analysis shows, however, that any deterministic local theories are necessarily incompatible with some other quantum-mechanical predictions for the *Gedankenexperiment*. Upon examining the proof in Bell's earlier paper, one might conjecture that it is essentially the deterministic character of the class of theories that is incompatible with quantum mechanics. That is, if the hypotheses assumed for the *Gedankenexperiment* were slightly relaxed so that determinism is no longer derivable, then the incompatibility with the quantum-mechanical predictions will also be removed. This conjecture is incorrect. Bell shows in his second paper that any stochastic theory satisfying the locality condition is also incompatible with quantum mechanics.

The conjecture that "it is essentially the deterministic character of the class of theories that is incompatible with quantum mechanics" is in fact inconsistent with the acknowledgement, expressed in the passage quoted above, that any local theory *must also be deterministic*, namely if you buy locality, *eo ipso* you buy determinism. Accordingly, the sentence "if the hypotheses assumed for the *Gedankenexperiment* were slightly relaxed so that determinism is no longer derivable, then the incompatibility with the quantum-mechanical predictions will also be removed" is awkward: since determinism, under the Bell 1964 conditions, *is* derivable, logically determinism *cannot* be the focus for assessing the compatibility or incompatibility of the local theory with quantum-mechanical predictions![23]

After this peculiar oscillation, CH go on to introduce a class of theories, called *objective local theories*, and to justify their approach as a true generalization with respect both to CHSH and Bell 1971 mainly under the heading of experimental feasibility: "Incompatibility of this class of theories with quantum mechanics has essentially been demonstrated by Bell, but the result is in a form that is not practically experimentally testable." (CH, p. 526, the reference to Bell is to the class of stochastic hidden variable theory introduced in Bell 1971). As to the term *objective*, what one can clearly gain from the paper is that the CH framework is designed to require as least as possible from the intuitive notion of physical 'object', not only in order to be more general than the existing approaches but also and above all because no 'determinism' is supposed to be available in their framework, no matter what is the role they take determinism to play – correctly or not – in the CHSH framework. In the paper section explicitly entitled "Objective local theories" (the section II), simply a source of coincident two-particle emissions is considered as usual (either with spin-1/2 particles or photons). During a selected period of time, where the adjustable parameters are denoted by $a$ and $b$, the source emits $N$ of the two-particle systems of interest. If $N_1(a)$ and $N_2(b)$ denote the number of counts at detectors 1 and 2, respectively, and if $N_{12}(a, b)$ denotes the number of coincident counts, it is possible to assume that, with $N$ sufficiently large, the ensemble probabilities of these results are the following

$$p_1(a) = N_1(a) / N$$
$$p_2(b) = N_2(b) / N$$

---

[23] As a matter of fact, CH refer the conjecture to Karl R. Popper and they do it in such a way that it is far from clear whether they endorse or not the idea that "upon examining the proof in Bell's earlier paper" it would be at least legitimate in principle to propose such conjecture. The conjecture is proposed in a collection in honor of Alfred Landé (Popper 1971), and Bell himself briefly dispenses with it in a review of the collection appeared in the issue of September 8th, 1972 on *Science* (Bell 1972).

$$p_{12}(a, b) = N_{12}(a, b) / N$$

If $\lambda$ is taken to denote "the state specification of the above system at a time intermediate between its emission and its impingement on either apparatus" (CH 1974, p. 527), it is natural to assume that $\lambda$ determines the single probabilities of counts at either apparatus, denoted by $p_1(\lambda, a)$ and $p_2(\lambda, b)$, and the coincidence count probability $p_{12}(\lambda, a, b)$. If, as customary, $\lambda$ is associated with a normalized probability density $\rho(\lambda)$, the probabilities $p_1(a)$, $p_2(b)$ and $p_{12}(a, b)$ become

$$p_1(a) = \int_\Lambda p_1(\lambda, a) \rho(\lambda) \, d\lambda$$

$$p_2(b) = \int_\Lambda p_2(\lambda, b) \rho(\lambda) \, d\lambda$$

$$p_{12}(a, b) \; p_{12}(\lambda, a, b) \; \int_\Lambda p_{12}(\lambda, a, b) \rho(\lambda) \, d\lambda$$

At this point, it is sufficient to impose a locality condition in terms of a factorization such that

$$p_{12}(\lambda, a, b) = p_1(\lambda, a) p_2(\lambda, b) \qquad (*)$$

in order to obtain an inequality (called the CH inequality) that is contradicted by the quantum-mechanical predictions. It is essential to recall that CH, *before* requiring locality as probability factorization, explicitly claim that the formulation "is quite general. *Nothing so far has been assumed that is not satisfied by quantum mechanics.*" (CH 1974, pp. 527-528). After motivating (*) in the usual way, to the effect that the space-like separation between the spacetime regions in which the triggering events take place justifies the probability factorization, CH state explicitly: "we call any theory in which [(*)] holds an objective local theory." (CH 1974, p. 528).

Still in his retrospect article we mentioned with respect to CHSH, Clauser seems to present the CH approach with motivations that cannot be found in the original 1974 paper, and it is interesting to analyze the mismatch between the two sources. In presenting the CH motivations, he writes:

> Mike and I thus asked ourselves the following questions: If determinism is not required of a system for Bell's Theorem's constraints to hold for it, what then is the fundamental characteristic of a physical system, such that it is constrained by Bell's Theorem? Also, how does one go about specifying said system's assumed nature within the Theorem's derivation? Our 1974 paper was the first publication that specifically answers these questions. (Clauser 2002, pp. 83-84).

In fleshing out the answers, Clauser gives a tone to the narrative that is distinctively more philosophically sounding than the sober, original presentation:

> Historically, the CHSH result can be seen as a logical continuation of a sequence of quandaries faced by Einstein during his development of special relativity. In formulating special relativity, Einstein noted that one cannot ask the universal questions "When" and "Where" in a precise fashion, without first defining the associated primitive entities, "time" and "distance". He found that, at best, he could only provide purely operational definitions for these entities. That is, he defines "time" to be the stuff you measure with a clock, and "distance" to be the stuff you measure with a ruler. Similarly, to allow Einstein to ask the universal question "What" in a similarly precise fashion, EPR's definition of an "element of reality" effectively provides his initial attempt to help one to define the associated primitive entity, "object". But, as noted above, that definition does not apply to realizable systems. CH thus found it necessary to offer an improved and explicit operational definition for the notion of an "object". The historical development of Bell's Theorem is then seen to have a direct parallel to the development of special relativity. Once that a suitably precise definition is given for some fundamental "stuff" of nature, then and only then can new testable physics emerge. (Clauser 2002, p. 84).

Two specific points deserve attention. First, Clauser claims that "EPR's definition of an 'element of reality' effectively provides his initial attempt to help one to define the associated primitive entity, 'object' ". That definition does not, however, help to define any 'primitive entity' or 'object': EPR take objects for granted and the EPR definition provides a condition for characterizing a *property*, that we have reason to attribute to an object even when we do not interact with it. Moreover, as stressed in the section 3, to assume such a definition *does not amount* to assume *ipso facto* the *existence* of the properties that the condition characterizes. Second, we can verify by inspection that the CH paper provides no "explicit operational definition for the notion of an 'object' ": the CH approach is based on an operational framework that, except for the probabilistic connection between the $\lambda$ and the detection at the apparatus, is completely similar in principle to the preceding Bell and CHSH papers (and others), and the paper presents no attempt at any sort of operationally founded ontology, as the above self-praising quotation by Clauser is meant to suggest. Therefore, there seems to be no ground in the CH paper to set the historical development of the Bell theorem and the development of special relativity in parallel, nor to acknowledge the relevance of any 'precise definition' for the 'stuff' of nature.

Moreover, due to the influence of a reading that in 2002 was already indisputable, Clauser 2002 presents the ghostly 'operational definition of an object' as consistent with 'local realism', even if – as we just have seen – the stochastic hidden variable theories' framework introduced in the CH paper does not assume, nor need, any 'realistic' assumption of any kind. The following passage shows how this misinterpretation affects even this exercise of view in retrospect:

> In their 1974 paper, CH operationally define an "object" (within a local realistic theory) as stuff with measurable properties that also can be put in a box, i.e. stuff (along with an associated measuring apparatus) that can be spatially surrounded by a space-time Gaussian surface. CH supplement this definition with an associated definition of an "objective local theory" […], as a theory that describes such objects, wherein action-at-a-distance is precluded, and wherein a measurement reveals local properties of an object (i.e. of said stuff within said box). Moreover, which properties of an object are to be measured may be arbitrarily chosen at the free will of the experimenter who operates the apparatus. Thus, for objects within such a theory, if the correlated properties of two objects in two disjoint boxes are measured at space-like separated measurement events, then absence of action-at-a-distance prevents the experimenter's choice of property measured in the first box (e.g. his choice of color vs. weight, in the earlier examples) from affecting the results of a measurement made on/in the second arbitrarily distant box. CH found that these very simple and naIve premises and definitions for *any* such local realistic theory of nature, then lead to a very general formulation of Bell's Theorem, that, in turn, then constrains the experimental predictions for said theory. The CH specific experimental predictions are contained in a new inequality that then must be satisfied by any such theory within local realism. That inequality is now commonly referred to as the CH inequality. (Clauser 2002, p. 84).

Not surprisingly, therefore, Clauser 2002 equates an 'objective local theory' of the CH 1974 paper with a 'local-realistic' theory, whereas the details of the CH 1974 paper do *not* support, or even require, such an equation, and evokes another influential review paper such as Clauser-Shimony 1978 as a reference paper: in the words of Clauser 2002, an objective local theory in the spirit of CH 1974 is "subsequently also called a "local realistic theory" and/or a "theory of local realism" by Clauser and Shimony" in their 1978 paper (Clauser, Shimony 1978).

This paper by Clauser and Shimony (CS from now on), a long and articulated review paper that thoroughly summarizes most of the work done about the Bell theorem both theoretically and experimentally up to 1978, can be considered an official and seemingly already undisputed celebration of the view according to which it is 'local realism' that the Bell theorem essentially challenges: the path from locality to 'local realism' appears to be over, so that the 'local-realistic' reading of the Bell theorem becomes folklore and common knowledge. In order to appreciate it, we recall that the paper opens with the very definition of *realism* and with the claim that a key issue in the foundations of quantum mechanics is exactly the search for consistency with *this* realism:

Realism is a philosophical view, according to which external reality is assumed to exist and have definite properties, whether or not they are observed by someone. So entrenched is this viewpoint in modern thinking that many scientists and philosophers have sought to devise conceptual foundations for quantum mechanics that are clearly consistent with it. (CS 1978, p. 1883)

According to CS, the locality problem addressed by the Bell theorem is formulated *within* this 'realistic framework'.

One possibility, it has been hoped, is to reinterpret quantum mechanics in terms of a statistical account of an underlying hidden-variables theory in order to bring it within the general framework of classical physics. However, Bell's theorem has recently shown that this cannot be done. The theorem proves that all realistic theories, satisfying a very simple and natural condition called locality, may be tested with a single experiment against quantum mechanics. These two alternatives necessarily lead to significantly different predictions. The theorem has thus inspired various experiments, most of which have yielded results in excellent agreement with quantum mechanics, but in disagreement with the family of local realistic theories. *Consequently, it can now be asserted with reasonable confidence that either the thesis of realism or that of locality must be abandoned. Either choice will drastically change our concepts of reality and of space-time.*

It is clear, then, that CS take for granted that locality *and* realism are both independent assumption of the (original) Bell theorem. But let us consider the CS reconstruction of the EPR argument, which immediately follows the above passage:

Within the realistic framework, Einstein *et al.* (1935, hereafter referred to as EPR) presented a classic argument. As a starting point, *they assumed the non-existence of action-at-a-distance* and that *some of the statistical predictions of quantum mechanics are correct*. They considered a system consisting of two spatially separated but quantum-mechanically correlated particles. For this system, *they showed that the results of various experiments are predetermined*, but that this fact is not part of the quantum-mechanical description of the associated systems. Hence that description is an incomplete one. To complete the description, it is thus necessary to postulate additional 'hidden variables', which presumably will then restore completeness, determinism and causality to the theory. (CS 1978, p. 1883).

Curiously, and once again, this (correct) reconstruction is inconsistent with the 'local-realistic' reading: the conjunction of locality ("*they assumed the non-existence of action-at-a-distance*") and the validity of selected quantum-mechanical predictions ("*some of the statistical predictions of quantum mechanics are correct*") is sufficient to derive the pre-determination ("*they **showed** that the results of various experiments are predetermined*"). As a consequence, this holds true also for the Bell theorem, which has the EPR argument as the main source. But few lines later a similar inconsistency emerges again:

EPR had demonstrated that any ideal system which satisfies a locality condition must be deterministic (at least with respect to the correlated properties). Since that argument applies only to ideal systems, CHSH therefore had postulated determinism explicitly. Yet, it eventually became clear that it is not the deterministic character of these theories that is incompatible with quantum mechanics. Although not stressed, this point was contained in Bell's subsequent papers (1971, 1972)-any non-deterministic (stochastic) theory satisfying a more general locality condition is also incompatible with quantum mechanics. Indeed it is the objectivity of the associated systems and their locality which produces the incompatibility. Thus, the whole realistic philosophy is in question!

Although in the first part of this passage CS acknowledge that (i) according to the EPR assumptions "any ideal system which satisfies a locality condition *must* be deterministic" – namely, determinism is a logical consequence of locality (plus quantum-mechanical predictions) – and (ii) the postulation of determinism in the CHSH framework was inessential, in the second part, in a surprising exercise of *non-sequitur* argumentation, CS go on to claim that the incompatibility is the result of locality *jointly with* an (ill-defined) assumption of 'objectivity', *in spite of* the irrelevance of determinism. The final slogan ("the whole realistic philosophy is in question!") puts then the tombstone on any attempt to distinguish between locality and 'realism' in the analysis of the puzzling consequences of the Bell theorem and its variants.

**Conclusions**

Nearly sixty years have elapsed from the first appearance of a theorem, due to J.S. Bell, that would have shaken the confidence in the possibility of a quiet coexistence of two pillars in the natural world as described by our best physics, quantum theory and special relativity. Seen in retrospect, the history of the debates on the foundational implications of the theorem displayed very soon an attempt to put the theorem in a perspective that was not entirely motivated by its very assumptions: the history itself of quantum theory appeared to be affected heavily by the issue of what the fate of a not-well-specified 'realism' would have been, so heavily that, already in the temporal surroundings of the Bell theorem publication, the interpretation of its general meaning could not avoid to fall into the intense gravitational field of the 'local-realistic' narrative. In an interview appeared in a collection devoted to leading physicists on issues of interpretation (Davies, Brown 1986), Davies asks Bell's opinion about *his* inequality presenting it as follows:

> Bell's inequality rooted in two assumptions: the first is what we might call objective reality – the reality of the external world, independent of our observations; the second is locality, or non-separability, or no faster-than-light signalling. Now, Aspect's experiment appears to indicate that one of these has to go. Which of the two would you like to hang on to?" (Davies, Brown 1986, p. 48) [24].

Interestingly, in his reply Bell focuses on Lorentz invariance and ignores altogether the assumption of 'objective reality', suggesting clearly – though implicitly – that experimentally confirmed violations of his inequality concern the former and not the latter (Davies, Brown 1986, pp. 48-49)[25].

---

[24] By the way, in the above text Davies appears to conflate conditions – locality, non-separability, no faster-than-light signalling – that are known to be *inequivalent*.

In the above pages, I attempted to focus on the complex and intricate steps through which the 'local-realistic' narrative set to establish as the current folklore already in the first fifteen years since the publication of the first version of the Bell theorem: the interpretive oscillations and inconsistencies that my analysis has put into focus, emerging in the very descriptions that many leading figures provided themselves of the deep work they devoted to the theorem and its consequences, show in a very illuminating way how the warning that the devil is in the details holds true *also* with respect to the issue of the meaning and import of the Bell theorem.

**Declaration of interest:** None.

---

[25] According to the above analysis, which has been supported by many (e.g. Norsen 2007, XXX 2008, 2012, 2018, Maudlin 2014, Tumulka 2016, and others), historical and textual evidence shows that already the original Bell 1964 contribution was quite clear that no 'local realism', but rather just locality, is at stake. Still, the issue is taken to be controversial by others (e.g. Werner 2014, Wiseman 2014, Griffiths 2020). In a recent contribution Hall argues that the form of locality used in the Bell 1964 is too weak for the derivation of pre-determination of values (Hall 2020). This argument assumes that the original 1964 Bell framework depends on a decomposition of locality proposed twenty years later, namely the decomposition into what have been called *parameter independence* and *outcome independence*. I argue that such a dependence cannot be defended on historical grounds: in fact, the Bell 1964 discussion takes locality as a *unitary* condition, not to mention that there are independent arguments to question the conceptual plausibility of such decomposition (see e.g. Norsen 2009, Maudlin 2011).